\documentclass[a4paper,12pt,reqno]{article}
\usepackage{amsmath}

\allowdisplaybreaks
\numberwithin{equation}{section}

\newcommand{\al}{\alpha}
\newcommand{\ad}{{\dot{\alpha}}}
\newcommand{\be}{\beta}
\newcommand{\bd}{{\dot{\beta}}}
\newcommand{\ep}{\varepsilon}
\newcommand{\la}{\lambda}
\newcommand{\bla}{\bar{\lambda}}
\newcommand{\si}{\sigma}
\newcommand{\bsi}{\bar{\sigma}}
\newcommand{\half}{\tfrac{1}{2}}
\newcommand{\ihalf}{\tfrac{\mathrm{i}}{2}}
\newcommand{\quart}{\tfrac{1}{4}}
\newcommand{\tab}{\quad\,}
\newcommand{\bD}{\bar{D}}
\newcommand{\hV}{\hat{V}}
\newcommand{\hF}{\hat{F}}
\newcommand{\p}{\partial} 

\newcommand{\na}[1]{\nabla_{\!#1}}	
\newcommand{\f}[2]{f_{#1}{}^{#2}}	
\newcommand{\T}[3]{T_{\!#1\,#3}^{#2}}	
\newcommand{\com}[2]{[\,#1\, ,\,#2\,]}	
\newcommand{\aco}[2]{\{#1\, ,\,#2\}}	

\newcommand{\cD}{\mathcal{D}}
\newcommand{\cDb}{\bar{\mathcal{D}}}
\newcommand{\cW}{\mathcal{W}}
\newcommand{\cH}{\mathcal{H}}
\newcommand{\cG}{\mathcal{G}}
\newcommand{\Ii}{\mathrm{i}}
\newcommand{\Ee}{\mathrm{e}}

\DeclareMathOperator{\re}{Re}
\DeclareMathOperator{\im}{Im}

\begin{document}

\begin{flushright}
 ITP-UH-26/98 \\ hep-th/9811180
\end{flushright}
\medskip

\begin{center}
 {\large\bfseries D=4, N=1 Supersymmetric Henneaux-Knaepen Models}
 \\[5mm]
 Friedemann Brandt and Ulrich Theis \\[2mm]
 \textit{Institut f\"ur Theoretische Physik, Universit\"at Hannover \\
	Appelstra\ss{}e 2, 30167 Hannover, Germany \\[2mm]
	e-mail: brandt@itp.uni-hannover.de, utheis@itp.uni-hannover.de}
\end{center}
\vspace{6mm}

\centerline{\bfseries Abstract} \medskip

We construct $N=1$ supersymmetric versions of four-dimensional
Freedman-Townsend models and generalizations thereof found recently by
Henneaux and Knaepen, with couplings between 1-form and 2-form gauge
potentials. The models are presented both in a superfield formulation
with linearly realized supersymmetry and in WZ gauged component form. In
the latter formulation the supersymmetry transformations are nonlinear
and do not commute with all the gauge transformations. Among others, our
construction yields $N=1$ counterparts of recently found $N=2$
supersymmetric gauge theories involving vector-tensor multiplets with
gauged central charge.

\parindent 1em

\section{Introduction}

Four-dimensional Freedman-Townsend models \cite{FT} involve peculiar
gauge invariant self-couplings of 2-form gauge potentials. These
couplings are local, but nonpolynomial in the fields and in the coupling
constant. Nonpolynomial couplings of a similar type, but between 2-form
gauge potentials and ordinary gauge fields, are met in recently
constructed $D=4$, $N=2$ supersymmetric gauge theories
\cite{CdWFKST,DT} involving so-called vector-tensor multiplets
\cite{SSW,dWKLL}.

In the latter models, the nonpolynomial couplings arise from gauging a
nonstandard global symmetry, the so-called central charge of the
vector-tensor multiplet. This was illustrated in \cite{BD} through a
nonsupersymmetric toy-model. In contrast, the nonpolynomial couplings
appearing in Freedman-Townsend models are not related to a global
symmetry that is gauged.

Nevertheless there is a relationship between all these models from which
one can also understand the origin and similarity of the peculiar
couplings appearing in them. In fact, the Freedman-Townsend models, the
toy-model discussed in \cite{BD}, and the purely bosonic part of the
$N=2$ supersymmetric gauge theories in \cite{CdWFKST,DT}, with scalars
set to constants, can all be fit in a larger class of
(nonsupersymmetric) gauge theories found recently by Henneaux and
Knaepen \cite{HK}.

As Henneaux-Knaepen models arise so naturally in the supersymmetric
gauge theories mentioned above, it is tempting to seek supersymmetric
versions of these models. The purpose of the present paper is the
construction of $N=1$ globally supersymmetric Henneaux-Knaepen models in
four spacetime dimensions.

We shall first review four-dimensional nonsupersymmetric
Henneaux-Knaepen models in section \ref{HK}. In section \ref{sfields} we
construct supersymmetric Henneaux-Knaepen models with linearly realized
supersymmetry in terms of appropriate superfields, generalizing earlier
work \cite{CLL} on supersymmetric Freedman-Townsend models. Section
\ref{WZ} provides the component version of these models in an
appropriate ``Wess-Zumino (WZ) gauge'' and is the main part of the
paper. In section \ref{example} we illustrate the results for two simple
examples, one of which is an $N=1$ counterpart of the aforementioned
$N=2$ gauge theories. The paper is ended with some concluding remarks in
section \ref{con} and a short appendix containing among others our
conventions.

\section{D=4 Henneaux-Knaepen models} \label{HK}

The models couple sets of 2-form and 1-form gauge
potentials. We shall label these gauge potentials by indices $A$ and
$a$ respectively, and denote their components by $B_{\mu\nu A} =
-B_{\nu\mu A}$ and $A_\mu^a$. The action and gauge transformations
can be elegantly written by means of auxiliary vector fields $V_\mu^A$.
In this first order formulation, they are polynomial. The nonpolynomial
form is then obtained upon eliminating the auxiliary fields. In first
order form, the Lagrangian reads
 \begin{align}
  L & = L_\mathrm{FT} + L_\mathrm{HK} + L_\mathrm{CM} + L_\mathrm{aux}
	\label{L} \\
  L_\mathrm{FT} & = -\quart\, \ep^{\mu\nu\rho\si} V_{\mu\nu}^A
	B_{\rho\si A} \label{LFT} \\
  L_\mathrm{HK} &= - \quart\, \delta_{ab} \hat{F}_{\mu\nu}^a
	\hat{F}^{\mu\nu b} \label{LHK} \\
  L_\mathrm{CM} & = -\tfrac{1}{8}\, c_{ab}\, \ep^{\mu\nu\rho\si}
	\hat{F}_{\mu\nu}^a \hat{F}_{\rho\si}^b \label{LCM} \\
  L_\mathrm{aux} & = \half\, \delta_{AB} V_\mu^A V^{\mu B} \label{Laux}
 \end{align}
where $c_{ab}$ are arbitrary real constant coefficients, and
$V_{\mu\nu}^A$ and $\hat{F}_{\mu\nu}^a$ are given by
 \begin{align}
  V_{\mu\nu}^A & = \p_\mu V_\nu^A - \p_\nu V_\mu^A + g \f{BC}{A}
	V_\mu^B	V_\nu^C \label{V} \\
  \hat{F}_{\mu\nu}^a & = \na{\mu} A_\nu^a - \na{\nu} A_\mu^a\ ,\quad
	\na{\mu} A_\nu^a = \p_\mu A_\nu^a + g V_\mu^A \T{A}{a}{b}
	A_\nu^b\ . \label{F}
 \end{align}
Here $g$ is a coupling constant with dimension $-1$, and $\f{BC}{A}$
and $\T{A}{a}{b}$ are real constants which satisfy
 \begin{gather}
  \f{[AB}{D} \f{C]D}{E} = 0 \label{jac} \\
  \T{A}{a}{c} \T{B}{c}{b} - \T{B}{a}{c} \T{A}{c}{b} = \f{AB}{C}
	\T{C}{a}{b}\ . \label{rep}
 \end{gather}
According to \eqref{jac} and \eqref{rep}, the $\f{AB}{C}$ are the
structure constants of a Lie algebra $\cG$, while the $\T{A}{a}{b}$ are
the entries of matrices $T_{\!A}$ representing $\cG$,
 \begin{equation*}
 \com{T_{\!A}}{T_{\!B}} = \f{AB}{C} T_{\!C}\ .
 \end{equation*}
Further conditions are not imposed.
In particular, $\cG$ can be any finite dimensional Lie algebra (not
necessarily compact), $T_A$ can be any real representation
thereof, and $\delta_{ab}$, $c_{ab}$ and $\delta_{AB}$,
which appear in $L_\mathrm{HK}$, $L_\mathrm{CM}$ and $L_\mathrm{aux}$
respectively, need not be $\cG$-invariant tensors (hence, in general $L$
is not globally $\cG$-invariant).

We shall refer to $L_\mathrm{FT}$, $L_\mathrm{HK}$, $L_\mathrm{CM}$ as
the Freedman-Townsend, Henneaux-Knaepen, and Chapline-Manton part of the
Lagrangian respectively. We note that by combining all these parts in a
single action we have slightly deviated from \cite{HK} where such a
combination was not considered (rather, Chapline-Manton type couplings,
of a more general form, were discussed separately from the other two
types). The reason is that the Chapline-Manton part arises naturally in
the supersymmetric extensions constructed later on, and therefore we
have introduced it already here. $L_\mathrm{CM}$ gives rise to couplings
of the 2-form gauge potentials (or, more precisely, their field
strengths) to Chern-Simons forms, similar to those appearing in
\cite{CM} and in the Green-Schwarz anomaly cancellation mechanism
\cite{GS}. This becomes clear upon elimination of the auxiliary fields
(see below). Note that the $V$-independent part of $L_\mathrm{CM}$ is a
total derivative.

Eq.\ \eqref{jac} guarantees the invariance of the action under the
following gauge transformations,
 \begin{equation} \label{gaugetrans}
  \delta_C B_{\mu\nu A} = \na{\mu} C_{\nu A} - \na{\nu} C_{\mu A}\
  ,\quad \delta_C A_\mu^a = 0\ ,\quad \delta_C V_\mu^A = 0\ ,
 \end{equation}
where the $C_{\mu A}$ are arbitrary fields, and $\na{\mu} C_{\nu A}$
is given by
 \begin{equation}\label{naC}
  \na{\mu} C_{\nu A} = \p_\mu C_{\nu A} - g V_\mu^B \f{BA}{C}
  C_{\nu C}\ .
 \end{equation}
Indeed, the $\delta_C$-transformation of the Lagrangian is a
total derivative,
 \begin{equation*}
  \delta_C L = \delta_C L_\mathrm{FT} = - \half\, \p_\rho
  (\ep^{\mu\nu\rho\si} V_{\mu\nu}^A C_{\si A})\ .
 \end{equation*}
This holds because the terms in $\delta_C L_\mathrm{FT}$ without
derivatives (i.e., those which are cubic in $V$) cancel thanks to
\eqref{jac}. One can easily deduce this from the Bianchi identity
 \begin{equation*}
  \ep^{\mu\nu\rho\si} \na{\rho} V_{\mu\nu}^A = \ep^{\mu\nu\rho\si}
  (\p_\rho V_{\mu\nu}^A + g V_\rho^B \f{BC}{A} V_{\mu\nu}^C) = 0\ .
 \end{equation*}
This Bianchi identity holds thanks to \eqref{jac} for any $\cG$, as
$V_{\mu\nu}^A$ has precisely the form of a nonabelian Yang-Mills
strength. Note however that $V_\mu^A$ cannot be interpreted as a
Yang-Mills gauge field (the action is clearly not invariant under
corresponding Yang-Mills gauge transformations, due to the presence of
$L_\mathrm{aux}$). The gauge transformations $\delta_C$ are
reducible because a shift $C_{\mu A} \rightarrow C_{\mu A} + \na{\mu}
Q_A$ modifies $\delta_C B_{\mu\nu A}$ only by the term
$\com{\na{\mu}}{\na{\nu}} Q_A = - gV_{\mu\nu}^B \f{BA}{C} Q_C$ which
vanishes on-shell for any fields $Q_A$ (as $V_{\mu\nu}^A$ vanishes 
by the equations of motion for $B_{\mu\nu A}$).

Eq.\ \eqref{rep} guarantees that the action is also gauge invariant
under
 \begin{equation} \label{split} \begin{split}
  \delta_\epsilon A_\mu^a & = \p_\mu \epsilon^a + g V_\mu^A \T{A}{a}{b}
	\epsilon^b \equiv \na{\mu} \epsilon^a\ ,\quad \delta_\epsilon
	V_\mu^A = 0\ , \\
  \delta_\epsilon B^{}_{\mu\nu A} & = g (\half\, \delta_{ab}\,
	\ep_{\mu\nu\rho\si} \hat{F}^{\rho\si a} - c_{ab}\,
	\hat{F}_{\mu\nu}^a) \T{A}{b}{c} \epsilon^c
 \end{split} \end{equation}
where the $\epsilon^a$ are arbitrary fields. Indeed, thanks to
\eqref{rep} one has
 \begin{equation*}
  \delta_\epsilon \hat{F}_{\mu\nu}^a = \com{\na{\mu}}{\na{\nu}}\,
  \epsilon^a = g V_{\mu\nu}^A\, \T{A}{a}{b}\, \epsilon^b\ .
 \end{equation*}
It is now easy to verify that $\delta_\epsilon (L_\mathrm{HK} +
L_\mathrm{CM})$ is precisely canceled by $\delta_\epsilon
L_\mathrm{FT}$,
 \begin{equation*}
  \delta_\epsilon (L_\mathrm{HK} + L_\mathrm{CM}) = - \delta_\epsilon
  L_\mathrm{FT} \quad \Rightarrow \quad \delta_\epsilon L = 0\ .
 \end{equation*}

Let us briefly discuss the formulation without auxiliary fields
$V_\mu^A$. Up to a total derivative, the Lagrangian takes the form
 \begin{equation*}
  \frac{1}{2}\, V_\mu^A\, K_{AB}^{\mu\nu}\, V_\nu^B - V_\mu^A
  \cH_A^\mu - \frac{1}{4}\, \delta_{ab} F_{\mu\nu}^a F^{\mu\nu b}\ ,
 \end{equation*}
where
 \begin{align*}
  K_{AB}^{\mu\nu} & = \delta_{AB} \eta^{\mu\nu} - \half g\, \f{AB}{C}
	\ep^{\mu\nu\rho\si} B_{\rho\si C} \\*
  & \tab - g^2 \T{A}{a}{c} \T{B}{b}{d} (\delta_{ab} \eta^{\mu\nu}
	A_\rho^c A^{\rho d} - \delta_{ab} A^{\mu d}\! A^{\nu c} -
	c_{ab}\, \ep^{\mu\nu\rho\si}\! A_\rho^c A_\si^d) \\[4pt]
  \cH_A^\mu & = \half \ep^{\mu\nu\rho\si} \p_\nu B_{\rho\si A}
	+ g \T{A}{a}{c} A_\nu^c \big( \delta_{ab} F^{\mu\nu b}
	+ \half c_{ab}\, \ep^{\mu\nu\rho\si} F_{\rho\si}^b \big)
	\\[4pt]
  F_{\mu\nu}^a & = \p_\mu A_\nu^a - \p_\nu A_\mu^a\ .
 \end{align*}
The equations of motion for $V_\mu^A$ give ($\approx$ denotes
equality on-shell)
 \begin{equation*}
  K_{AB}^{\mu\nu}\, V_\nu^B \approx \cH_A^\mu \quad \Rightarrow \quad
  V_\mu^A \approx (K^{-1})_{\mu\nu}^{AB} \cH_B^\nu\ ,\quad
  (K^{-1})_{\mu\rho}^{AC}\, K_{CB}^{\rho\nu} = \delta_B^A\,
  \delta_\mu^\nu\ .
 \end{equation*}

The formulation without auxiliary fields is thus obtained by
substituting $(K^{-1})_{\mu\nu}^{AB} \cH_B^\nu$ for $V_\mu^A$ in the
above expressions for the action and gauge transformations. For
instance, the action turns into
 \begin{equation} \label{action}
  S = - \int\! d^4x\, \big[ \half \cH^\mu_A\, (K^{-1})^{AB}_{\mu\nu}\,
  \cH^\nu_B + \quart \delta_{ab} F_{\mu\nu}^a F^{\mu\nu b} \big]\ .
 \end{equation}
Note that $K_{AB}^{\mu\nu}$ depends on the fields, but not on
derivatives thereof. Hence, its inverse is nonpolynomial in the fields,
but still local. As a consequence, in the formulation without auxiliary
fields, the action and gauge transformations are also
nonpolynomial but remain local. In fact, the action contains only terms
with exactly two spacetime derivatives, while the gauge transformations
are linear in derivatives. The gauge transformations commute on-shell,
i.e., they are abelian.

To understand the nature of the above models and of their gauge
symmetries, it is instructive to view them as deformations of
corresponding free theories (in fact, this is how they were derived in
\cite{HK}). The free action ($g=0$) reads
 \begin{equation*}
  S_\mathrm{free} = - \int\! d^4x\, \big[ \half H_\mu^A H^\mu_A
  +\quart F_{\mu\nu}^a F^{\mu\nu}_a \big]
 \end{equation*}
where
 \begin{equation*}
  H^\mu_A = \half \ep^{\mu\nu\rho\si} \p_\nu B_{\rho\si A}\ ,\quad
  H_\mu^A = \delta^{AB} H_{\mu B}\ ,\quad F^{\mu\nu}_a = \delta_{ab}
  F^{\mu\nu b}\ .
 \end{equation*}
The free theory has, among others, global symmetries generated by
 \begin{equation*}
  \Delta_a A_\mu^b = H_\mu^A\, \T{A}{b}{a}\, ,\quad \Delta_a
  B_{\mu\nu A} = \half\, \ep_{\mu\nu\rho\si}
  F^{\rho\sigma}_b\, \T{A}{b}{a}
 \end{equation*}
and corresponding Noether currents
 \begin{equation*}
  j_a^\mu = \T{A}{b}{a}\, H_\nu^A\, F^{\mu\nu}_b\ .
 \end{equation*}
Furthermore, it possesses conserved nontrivial currents of second
order,
 \begin{equation*}
  j^{\mu\nu A} = \half\, \f{BC}{A} \ep^{\mu\nu\rho\si} H_\rho^B
  H_\sigma^C\ .
 \end{equation*}
Expanding the action \eqref{action} in $g$, one finds
 \begin{equation*}
  S = S_\mathrm{free} - g \int\! d^4x\, \big( B_{\mu\nu A}\,
  j^{\mu\nu A} + A_\mu^a\, j_a^\mu + c_{ac} \T{A}{c}{b}\, 
  \ep^{\mu\nu\rho\si}
  H_\mu^A A_\nu^a \p^{}_\rho A_\si^b \big) + O(g^2)\ .
 \end{equation*}
Hence, to first order in $g$ the action couples $A_\mu^a$ and
$B_{\mu\nu A}$ to first and second order currents $j_a^\mu$ and
$j^{\mu\nu A}$ of the free theory. These couplings arise from
$L_\mathrm{HK}$ and $L_\mathrm{FT}$ respectively. In addition
there are couplings of $H^A_\mu$ to abelian Chern-Simons forms
originating from $L_\mathrm{CM}$.

Analogously one may expand the gauge transformations in $g$. At zeroth
order this reproduces of course the gauge symmetries of the free theory.
The first order pieces involve the global symmetries of the free action
given above through transformations $g\epsilon^a\Delta_a$. Hence,
$\delta_\epsilon$ gauges these global symmetries ($\delta_\epsilon
B_{\mu\nu A}$ involves in addition terms related to the Chapline-Manton
couplings). This explains why Henneaux-Knaepen models arise when one
gauges the central charge of the $N=2$ vector tensor multiplet, as this
central charge is a global symmetry of the above type.

We remark that $j_a^\mu$, $j^{\mu\nu A}$ are conserved for any constants
$\T{A}{b}{a}$, $\f{BC}{A}$, i.e., whether or not these constants fulfill
\eqref{jac} and \eqref{rep}. The latter conditions arise at order $g^2$
from the requirement that the deformed action be invariant under deformed
versions of the gauge transformations of the free model \cite{HK}.

\section{Superfield formulation} \label{sfields}

We shall now construct a supersymmetric extension of the Lagrangian
\eqref{L} in terms of superspace integrals. To this end we associate an
appropriate superfield with each of the fields $B_{\mu\nu A}$, $A_\mu^a$
and $V_\mu^A$, and generalize the gauge transformations
\eqref{gaugetrans} and \eqref{split} to these superfields. Similarly to
the nonsupersymmetric case, the superfield associated with $V_\mu^A$ is
auxiliary and may eventually be eliminated algebraically. Our
construction applies to general $\cG$ (not necessarily compact), and any
real representation $T_A$ thereof. As a consequence, even in the pure
Freedman-Townsend case, in general we cannot use traces over matrix
valued fields in order to construct the Lagrangian. Therefore we must
spell out indices $A$ and $a$ explicitly, where necessary.

The superfields associated with $A_\mu^a$ and $V_\mu^A$ are standard
real vector superfields which we denote by $A^a$ and $V^A$ respectively.
We assign dimension 0 to $A^a$ and dimension 1 to $V^A$, as the latter
is auxiliary.

The superfield associated with $B_{\mu\nu A}$ is a spinorial one
as in \cite{Siegel} and denoted by $\Psi^\al_A$.
It is chiral,
 \begin{equation}
  \bD_\ad \Psi^\al_A = 0\ ,
 \end{equation}
where $\bD_\ad$ is a supercovariant derivative mapping superfields
into superfields, cf.\ appendix. We assign dimension $1/2$ to
$\Psi^\al_A$. Then the independent component fields of $\Psi^\al_A$ are
two Weyl fermions with dimension $1/2$ and $3/2$ respectively, a complex
scalar field with dimension 1 and a 2-form gauge potential, also with
dimension 1. We remark that $\Psi^\al_A$ is the prepotential of a real
linear superfield $\Phi_A = D^\al \Psi_{\al A} + \bD_\ad
\bar{\Psi}^\ad_A$ ($D^2 \Phi_A = 0$).

In order to construct the superfield action, we define two chiral
superfields, $Y^a_\al$ and $W^A_\al$, constructed of $A^a$ and $V^A$.
$Y_\al^a$ is given by
 \begin{equation} \label{Y}
  Y_\al^a = - \frac{\Ii}{4} \bD^2 (\Ee^{-2\Ii V} D_\al\,
  \Ee^{\Ii V}\!\! A)^a\ ,\quad V = g V^A T_A\ ,
 \end{equation}
where $T_A$ are real matrices representing $\cG$ as in \eqref{rep},
and $\bD^2 = \bD_\ad \bD^\ad$.
In \eqref{Y}, $\bD_\ad$ and $D_\al$ act on everything to their
right, and ordinary matrix multiplication is understood, i.e.,
 \begin{equation*}
  (\Ee^{-2\Ii V}\! D_\al\, \Ee^{\Ii V}\!\! A)^a = (\Ee^{-2\Ii V}
  )^a{}_bD_\al\, \big[ (\Ee^{\Ii V})^b{}_c A^c \big]\ .
 \end{equation*}
$W_\al^A$ is defined analogously to the spinorial field strength
in super-Yang-Mills theory,
 \begin{equation} \label{W}
  g W_\al^A\, T_A = - \frac{\Ii}{4} \bD^2 (\Ee^{-2\Ii V}\! D_\al\,
  \Ee^{2\Ii V})\ .
 \end{equation}

We are now prepared to present our superfield Lagrangian. It reads
 \begin{align}
  L & = L_\mathrm{FT} + L_\mathrm{HK} + L_\mathrm{CM}
        + L_\mathrm{FI} + L_\mathrm{aux}
	\label{sL} \\
  & L_\mathrm{FT} = - \frac{1}{2} \int\! d^2 \theta\, W^A \Psi_A
	+ \text{c.c.} \label{sFT} \\
  & L_\mathrm{HK} + L_\mathrm{CM} = \int\! d^2 \theta\, k_{ab} Y^a Y^b
	+ \text{c.c.} \label{sHK} \\
  & L_\mathrm{FI} = \int\! d^2 \theta\, d^2 \bar{\theta}\,
        \mu_{\hat{a}} A^{\hat{a}} \label{sFI} \\
  & L_\mathrm{aux} = \int\! d^2 \theta\, d^2 \bar{\theta}\,
	\mathcal{F}(V) + \text{c.c.} \label{saux} 
 \end{align}
where $k_{ab} = k_{ba}$ are arbitrary complex numbers, and
$\mathcal{F}(V)$ is any function of the $V^A$ that allows eventually to
eliminate $V^A$ (e.g., one may take $\mathcal{F}(V) \propto \delta_{AB}
V^A V^B$, but more general choices are admissible too). $L_\mathrm{FI}$
is present only in the special case that all the representation matrices
$T_A$ have a vanishing row in common, i.e., only if
 \begin{equation} \label{FIcond}
  \T{A}{\hat{a}}{b} = 0 \quad \forall A,b
 \end{equation}
for some $\hat{a}$. In that case one may include $L_\mathrm{FI}$, with
arbitrary real numbers $\mu_{\hat{a}}$. $L_\mathrm{FI}$ is of course a
Fayet-Iliopoulos contribution \cite{FI}. The supersymmetric
Henneaux-Knaepen and Chapline-Manton parts of the Lagrangian,
$L_\mathrm{HK}$ and $L_\mathrm{CM}$, arise from the real and imaginary
part of $k_{ab}$ respectively.

Thanks to the use of superspace techniques, the action $\int\! d^4x\,
L$ is manifestly supersymmetric. We shall now show that it has in
addition gauge symmetries corresponding to \eqref{gaugetrans} and
\eqref{split}. As in \cite{CLL},
the counterpart of \eqref{gaugetrans} is generated on the
superfields through
 \begin{equation} \label{dpsi}
  \delta_C \Psi_{\al A} = \Ii \bD^2 (\Ee^{-2\Ii\hV}\! D_\al\,
  \Ee^{\Ii\hV}\! C)_A\ ,\quad \delta_C A^a = 0\ ,\quad \delta_C
  V^A = 0
 \end{equation}
where the $C_A$ are arbitrary real vector superfields, and
 \begin{equation}
  \hV_A{}^B \equiv - g V^C \f{CA}{B}\ .
 \end{equation}
In \eqref{dpsi}, matrix multiplication is understood, as in
\eqref{Y}. In order to verify the invariance of the action, one
calculates
 \begin{align*}
  \delta_C\, L = \delta_C\, L_\mathrm{FT} & = - \frac{\Ii}{2}
	\int\! d^2 \theta\, W^{\al A} \bD^2 (\Ee^{-2\Ii\hV}\! D_\al\,
	\Ee^{\Ii\hV} C)_A + \text{c.c.} \\
  & \simeq 2\Ii \int\! d^2 \theta\, d^2 \bar{\theta}\, (W^\al
	\Ee^{-2\Ii\hV})^A D_\al\, (\Ee^{\Ii\hV} C)_A + \text{c.c.} \\
  & \simeq -2\Ii \int\! d^2 \theta\, d^2 \bar{\theta}\;  C_B\,
	\big[ (\Ee^{\Ii\hV})_A{}^B\, D^\al (W_\al\, \Ee^{-2\Ii\hV})^A
	- \text{c.c.} \big] \\
  & = 0\ ,
 \end{align*}
where $\simeq$ denotes equality up to total derivatives and we have
used that $C_A$ is real. The last equality holds thanks to the identity
 \begin{equation} \label{bianchi2}
  (\Ee^{\Ii\hV})_A{}^B\, D^\al (W_\al\, \Ee^{-2\Ii\hV})^A =
   (\Ee^{-\Ii\hV})_A{}^B\, \bD_\ad (\bar{W}^\ad \Ee^{2\Ii\hV})^A
 \end{equation}
where 
 \begin{equation*}
  (W_\al\, \Ee^{-2\Ii\hV})^A = W^B_\al\, (\Ee^{-2\Ii\hV})_B{}^A\ .
 \end{equation*}
\eqref{bianchi2} is nothing but the
``super-Bianchi identity'' (for any $\cG$) familiar from
super-Yang-Mills theory, cf.\ appendix.

\eqref{dpsi} extends indeed the gauge transformation \eqref{gaugetrans}
to the superfields, as the field $C_{\mu A}$ which appears in
\eqref{gaugetrans} corresponds just to the vector field contained in
$C_A$.

Finally we present the superfield version of the gauge transformations
\eqref{split}. It reads
 \begin{equation} \begin{split} \label{dA}
  \delta_\Lambda A^a & = \Ii (\Ee^{\Ii V}\! \Lambda - \Ee^{-\Ii V}
	\bar{\Lambda})^a\ ,\quad \delta_\Lambda V^A = 0 \\
  \delta_\Lambda \Psi_A & = 4\Ii g\, k_{ab} Y^a\, \T{A}{b}{c}\,
  \Lambda^c
 \end{split} \end{equation}
where the $\Lambda^a$ are abitrary chiral superfields,
 \begin{equation} \label{chir}
  \bD_\ad \Lambda^a = 0\ .
 \end{equation}
Using \eqref{chir}, one verifies that \eqref{dA} implies
 \begin{equation}
  \delta_\Lambda Y^a_\al = \Ii g\, W^A_\al\, \T{A}{a}{b}
  \Lambda^b\ .
 \end{equation}
It is now easy to check that the superfield Lagrangian is
$\delta_\Lambda$-invariant. Indeed, $L_\mathrm{aux}$ is evidently
invariant, while the transformations of $L_\mathrm{FT}$ and
$L_\mathrm{HK} + L_\mathrm{CM}$ cancel,
 \begin{equation}
  \delta_\Lambda\, (L_\mathrm{FT} + L_\mathrm{HK} + L_\mathrm{CM})
  = \int\! d^2 \theta\, (-\half W^A \delta_\Lambda \Psi_A + 2 k_{ab}
  Y^a \delta_\Lambda Y^b) + \text{c.c.} = 0\ .
 \end{equation}
Finally, if \eqref{FIcond} holds, then $\exp(\pm\Ii V)^{\hat a}{}_b
=\delta^{\hat a}_b$, and \eqref{split} implies
 \begin{equation*}
  \delta_\Lambda A^{\hat{a}} = \Ii (\Lambda^{\hat{a}} -
  \bar{\Lambda}^{\hat{a}})
 \end{equation*}
which in turn guarantees the gauge invariance of \eqref{sFI}, as
$\Lambda^{\hat{a}}$ is a chiral superfield. Note that $A^{\hat{a}}$
transforms exactly as a standard abelian gauge superfield.

The lowest component field of $\Lambda^a + \bar{\Lambda}^a$ corresponds
to $\epsilon^a$ in \eqref{split}.

\section{Models in WZ gauge} \label{WZ}

The gauge transformations \eqref{dpsi} and \eqref{dA} act as shift
symmetries on some of the component fields of the superfields $\Psi_A$
and $A^a$. As usual, this signals that the action can actually be
written in terms of fewer fields, with a correspondingly reduced gauge
invariance and modified supersymmetry transformations. In this section
we shall construct such a ``WZ gauged'' version of the models.

\eqref{dpsi} suggests that, in WZ gauge, the remaining fields
originating from $\Psi_A$ will be those of a real linear multiplet,
i.e., a real scalar field $\varphi_A$ with dimension 1, the components
$B_{\mu\nu A}$ of a real 2-form gauge potential, also with dimension 1,
and a Weyl spinor $\chi_A$ with dimension $3/2$. Similarly \eqref{dA}
indicates that, in WZ gauge, $A^a$ will give rise only to a real vector
field $A_\mu^a$ with dimension 1, a Weyl spinor $\lambda^a$ with
dimension $3/2$ and a real auxiliary field $D^a$ with dimension 2.

We shall now work in component formalism with such a field content
($B_{\mu\nu A}$, $\varphi_A$, $\chi_A$, $A_\mu^a$, $\lambda^a$, $D^a$).
Again, we complement these fields by all the component fields of the
auxiliary superfields $V^A$ in order to work in a convenient first order
formulation. This is possible because the latter fields are invariant
under the gauge transformations \eqref{dpsi} and \eqref{dA} and can thus
be kept in WZ gauge. As before, the component fields of $V^A$ are
auxiliary and may be eliminated algebraically at the end, along with the
$D^a$. Hence, in a formulation without auxiliary fields, one is left
with the field content $B_{\mu\nu A}$, $\varphi_A$, $\chi_A$, $A_\mu^a$,
$\lambda^a$.

Now, from the experience with other supersymmetric gauge theories, one
expects that the supersymmetry algebra holds in WZ gauge only modulo
gauge transformations. This is our motivation for using a particular
gauge covariant graded commutator algebra of supersymmetry and gauge
transformations as the starting point for the construction of WZ gauged
models. We shall then use this algebra to construct the 
Lagrangian, supersymmetry and gauge transformations.

The algebra has an unusual form which is inspired by the models in
sections \ref{HK} and \ref{sfields} (see discussion below). On gauge
covariant quantities constructed of $A_\mu^a$, $\lambda^a$, $D^a$ and
the component fields of $V^A$ it reads\footnote{One would not expect
that \eqref{algebra} can be realized also on $B_{\mu\nu A}$, $\varphi_A$
and $\chi_A$ since it does not contain the gauge transformations
\eqref{gaugetrans} and since the gauge transformations \eqref{split} do
not commute off-shell on $B_{\mu\nu A}$. Indeed, we shall find that the
algebra has an accordingly modified form on $B_{\mu\nu A}$, $\varphi_A$
and $\chi_A$.}

 \begin{align} \label{algebra}
  \com{\cD_\mu}{\cD_\nu}  & = - g \hat{F}_{\mu\nu}^a\, \delta_a &
	\com{\delta_a}{\cD_\mu} & = - g V_\mu^A \T{A}{b}{a}\, \delta_b
	\notag \\*
  \com{\cD_{\al}}{\cD_{\be\bd}} & = - 2g\, \ep_{\al\be}\,
	\bla^a_{\bd}\, \delta_a & \com{\delta_a}{\cD_\al} & = - g
	\Gamma^A_{\!\al} \T{A}{b}{a}\, \delta_b \notag \\
  \aco{\cD_\al}{\cD_\be} & = 0 & \com{\delta_a}{\delta_b} & = 0
	\notag \\
  \aco{\cD_\al}{\cDb_\ad} & = - \Ii\, \cD_{\al\ad}
 \end{align}
where $\cD_\al$ and $\cDb_\ad$ generate the supersymmetry
transformations (on component fields), the $\delta_a$ generate gauge
transformations corresponding to \eqref{split} resp.\ \eqref{dA},
$\Gamma^A$ and $V_\mu^A$ will be constructed of the auxiliary fields
(see below), and $\cD_\mu$ are gauge covariant derivatives
 \begin{equation} \label{D}
  \cD_\mu = \p_\mu - g A_\mu^a\, \delta_a\ .
 \end{equation}

Note that \eqref{algebra} is somewhat similar to the gauge covariant
algebra in WZ gauged super-Yang-Mills theories. However there is a
remarkable difference to the latter theories (and to other
supersymmetric gauge theories as well): the supersymmetry
transformations do not commute with all the gauge
transformations!\footnote{Actually the algebra \eqref{algebra} alone
would still permit the possibility that supersymmetry and gauge
transformations commute on-shell. However, this will not be the case, 
as one expects since
the gauge transformations \eqref{split} do not vanish
on-shell (cf.\ also
remarks at the end of this section).} In order to explain this 
unusual feature we remark:

\begin{enumerate}

\item[(a)] From sections \ref{HK} and \ref{sfields} it is clear that the
algebra \eqref{algebra} should in the special case $T_A =0$ reproduce
the supersymmetry algebra of usual abelian gauge theory in WZ gauge.
Hence, $\aco{\cD_\al}{\cDb_\ad}$ should thus contain the covariant
derivative rather than the partial one, reflecting the presence of a
gauge transformation in the commutator of two supersymmetry
transformations.

\item[(b)] We aim at the construction of supersymmetrized
Henneaux-Knaepen models. To that end $\com{\delta_a}{\cD_\mu}$ must not
vanish because otherwise we would get $\hat{F}_{\mu\nu}^a = \p_\mu
A_\nu-\p_\nu A_\mu$ rather than an expression like \eqref{F}. This is
seen from the following calculation which uses \eqref{D} in the form
$\p_\mu = \cD_\mu + g A_\mu^a\, \delta_a$:
 \begin{align*}
  0 & = \com{\p_\mu}{\p_\nu} = \p_\mu(\cD_\nu + g A_\nu^a\, \delta_a)
	- (\mu \leftrightarrow \nu) \\
    & = (\cD_\mu + g A_\mu^a\, \delta_a) \cD_\nu + g (\p_\mu A_\nu^a)\,
	\delta_a + g A_\nu^a\, (\cD_\mu + g A_\mu^b\, \delta_b)\,
	\delta_a - (\mu \leftrightarrow \nu) \\
    & = \com{\cD_\mu}{\cD_\nu} + g A_\mu^a\, \com{\delta_a}{\cD_\nu}
	- g A_\nu^a\, \com{\delta_a}{\cD_\mu} \\
    & \tab + g (\p_\mu A_\nu^a - \p_\nu A_\mu^a)\, \delta_a + g^2
	A_\nu^a	A_\mu^b\, \com{\delta_b}{\delta_a}\ .
  \end{align*}

\end{enumerate}

Since $\aco{\cD_\al}{\cDb_\ad} = - \Ii\, \cD_{\al\ad}$ and
$\com{\delta_a}{\cD_\al}=0$ would imply $\com{\delta_a}{\cD_\mu}=0$, the
requirements in (a) and (b) show that $\com{\delta_a}{\cD_\al}$ must not
vanish because otherwise we would not end up with supersymmetric
Henneaux-Knaepen models (rather, we would get a supersymmetric abelian
gauge theory of the standard type). Besides, the calculation in (b) also
shows that the algebra \eqref{algebra} reproduces exactly the curvature
\eqref{F},
 \begin{equation*}
  \hat{F}_{\mu\nu}^a = \p_\mu A_\nu^a - \p_\nu A_\mu^a + g \T{A}{a}{b}
  (V_\mu^A A_\nu^b - V_\nu^A A_\mu^b)\ .
 \end{equation*}

Now, an analysis of \eqref{algebra} and the Bianchi identities following
from it shows that the algebra is realized off-shell by the following
supersymmetry and gauge transformations of $A_\mu^a$, $\lambda^a$, $D^a$
and the component fields of $V^A$:

\begin{enumerate}

\item[(i)] All the component fields of $V^A$ are gauge invariant and
\eqref{algebra} reduces thus on these fields to the standard
supersymmetry algebra. Hence, the component fields of $V^A$ form a
standard real $N=1$ vector multiplet, as in the superfield formulation.
$\Gamma^A_{\!\al}$ is defined through
 \begin{equation} \label{Gamma}
  g \Gamma^A_{\!\al}\, T_A = (\Ee^{-\Ii V}\! D_\al\, \Ee^{\Ii V})|\
  ,\quad g \bar{\Gamma}^A_{\!\ad}\, T_A = (\Ee^{\Ii V}\! \bar D_\ad\,
  \Ee^{-\Ii V})|
  \end{equation}
where $|$ denotes the $\theta$-independent part of a superfield,
and we used a notation as in \eqref{W}. This implies
 \begin{equation*}
  \cD^{}_\al \Gamma^A_{\!\be} + \cD^{}_\be \Gamma^A_{\!\al} + g
  \Gamma^B_{\!\al} \Gamma^C_{\!\be} \f{BC}{A} = 0\ ,
 \end{equation*}
as required by the Bianchi identity 
 \begin{equation*}
  \aco{\com{\delta_a}{\cD_\al}}{\cD_\be} +
  \com{\aco{\cD_\al}{\cD_\be}}{\delta_a} -
  \aco{\com{\cD_\be}{\delta_a}}{\cD_\al} = 0\ .
 \end{equation*}
The analogous Bianchi identity with $\cD_\be$ replaced by
$\cDb_\ad$ determines $V_\mu^A$,
 \begin{equation}
  V_{\al\ad}^A = \Ii (\cD^{}_\al \bar{\Gamma}^A_{\!\ad} +
  \cDb^{}_\ad \Gamma^A_{\!\al} + g \f{BC}{A} \Gamma^B_{\!\al}
  \bar{\Gamma}^C_{\!\ad})\ .
 \end{equation}

For later purpose we note that one gets
\begin{align}
 \cD_\al V_\mu^A & = - (\si_\mu \bar{\eta}^A)_\al + \p_\mu
	\Gamma^A_{\!\al} - g \Gamma^B_{\!\al} V_\mu^C \f{BC}{A}
	\notag \\[4pt]
 \cD_\al \eta_\be^A & = \ep_{\al\be} h^A + \ihalf
	\si^{\mu\nu}{}_{\!\!\al\be} V_{\mu\nu}^A - g \Gamma^B_{\!\al}
	\eta_\be^C \f{BC}{A} \notag \\[4pt]
 \cDb_\ad \eta_\al^A & = - g \bar{\Gamma}^B_{\!\ad} \eta_\al^C
	\f{BC}{A} \notag \\[4pt]
 \cD_\al h^A & = -\ihalf \p_{\al\ad} \bar{\eta}^{A \ad} - \ihalf g
	V_{\al\ad}^B \bar{\eta}^{C\ad} \f{BC}{A} - g \Gamma^B_{\!\al}
	h^C \f{BC}{A} \notag \\[4pt]
 \delta_a \eta_\al^A & = \delta_a V_\mu^A = \delta_a h^A = 0\ ,
	\label{DV}
 \end{align}
where $\eta^A$, $h^A$ and $V_{\mu\nu}^A$ can be obtained from a
superfield $\cW^A$ defined in the appendix,
 \begin{align}
  \eta_\al^A & = \cW_\al^A| \notag \\[4pt]
  h^A & = \quart (D^\al \cW^A_\al + g \f{BC}{A} \Gamma^{\al B}
	\cW^C_\al)| + \text{c.c.}\ , \notag \\[4pt]
  V_{\mu\nu}^A & = -\Ii \si_{\mu\nu}{}^{\al\be} (D_\al
	\cW^A_\be + g \f{BC}{A} \Gamma^B_{\!\al} \cW^C_\be)|
	+ \text{c.c.} \notag \\
  & = \p_\mu V_\nu^A - \p_\nu V_\mu^A + g V_\mu^B V_\nu^C \f{BC}{A}\ .
	\label{curved}
 \end{align}

\item[(ii)] The supersymmetry transformations of $A_\mu^a$,
$\la^a_\al$ and $D^a$ are given by
 \begin{align}
  \cD_\al A_\mu^a & = - (\si_\mu \bla^a)_\al - g \Gamma^A_{\!\al}
	\T{A}{a}{b}\, A_\mu^b \notag \\[4pt]
  \cD_\al \la_\be^a & = \ep_{\al\be} D^a + \ihalf
	\si^{\mu\nu}{}_{\!\!\al\be} \hat{F}_{\mu\nu}^a - g
	\Gamma^A_{\!\al} \T{A}{a}{b}\, \la_\be^b \notag \\[4pt]
  \cDb_\ad\la_\al^a & = - g \bar{\Gamma}^A_{\!\ad} \T{A}{a}{b}\,
	\la_\al^b \notag \\[4pt]
  \cD_\al D^a & = - \ihalf (\cD_\mu + g V_\mu^A T_A)^a{}_b\,
	(\si^\mu \bla^b)_\al - g \Gamma^A_{\!\al} \T{A}{a}{b}\,
	D^b\ . \label{DA}
 \end{align}
$\delta_a$ is realized on $\la^b$, $\hat{F}_{\mu\nu}^b$ and $D^b$ by
 \begin{equation} \label{gauge_gauge}
  \delta_a \la_\al^b = \eta_\al^A \T{A}{b}{a}\, ,
  \quad \delta_a \hat{F}_{\mu\nu}^b = V_{\mu\nu}^A \T{A}{b}{a}\, 
  \quad \delta_a D^b = h^A \T{A}{b}{a}\ .
 \end{equation}
The corresponding gauge transformations of $A_\mu^b$, $\la^b$
and $D^b$ are
 \begin{equation} \label{gauge_gauge1}
  \delta_\epsilon A_\mu^b = \p_\mu \epsilon^b + g V_\mu^A \T{A}{b}{a}
  \epsilon^a\ ,\quad \delta_\epsilon \la_\al^b = g\, \eta_\al^A
  \T{A}{b}{a} \epsilon^a\ , \quad \delta_\epsilon D^b = g\, h^A
  \T{A}{b}{a} \epsilon^a\ .
 \end{equation}

\end{enumerate}

We are now prepared to construct the WZ gauged Lagrangian, along with
the supersymmetry and gauge transformations of $B_{\mu\nu A}$,
$\varphi_A$ and $\chi_A$. The Lagrangian is
 \begin{align}
  & L = L_\mathrm{FT} + L_\mathrm{HK} + L_\mathrm{CM} + L_\mathrm{FI}
	+ L_\mathrm{aux} \\
  & L_\mathrm{FT} = - \frac{1}{4}\, \ep^{\mu\nu\rho\si}\, V_{\mu\nu}^A\,
	B_{\rho\si A} + h^A \varphi_A + \eta^A \chi_A + \bar{\eta}^A
	\bar{\chi}_A \label{WZFT} \\
  & L_\mathrm{HK} + L_\mathrm{CM} = - \frac{1}{4}\, \cD^2 \big[
	k_{ab}\, (\Ee^{-\Ii v} \lambda)^a (\Ee^{-\Ii v} \lambda)^b
	\big] + \text{c.c.} \label{WZHK} \\
  & L_\mathrm{FI} = \mu_{\hat{a}} D^{\hat{a}} \label{WZFI}
 \end{align}
with $L_\mathrm{aux}$ as in \eqref{saux} (since the component fields of
$V^A$ are gauge invariant and have the same supersymmetry
transformations as in the superfield formulation). In \eqref{WZHK} we
used the notation $\cD^2=\cD^\al \cD_\al$, $k_{ab}$ are abitrary complex
numbers as in \eqref{sHK}, and $v$ is a matrix valued field constructed
of the lowest component fields of the $V^A$ and the representation
matrices $T_A$,
 \begin{equation}
  (\Ee^{-\Ii v} \lambda_\al)^a = (\Ee^{-\Ii v})^a{}_b\, \lambda_\al^b\
  ,\quad v = g v^A\, T_A\ ,\quad v^A = V^A|\ .
 \end{equation}
\eqref{WZHK} will be spelled out explicitly at the end of this section.

As in the superfield formulation, the Fayet-Iliopoulos part \eqref{WZFI}
is present only if all the representation matrices $T_A$ have a
vanishing row in common, i.e., if \eqref{FIcond} holds. $L_\mathrm{aux}$
and $L_\mathrm{FI}$ are separately supersymmetric (up to total
derivatives) and gauge invariant and therefore need not be discussed
further (indeed, \eqref{FIcond}, \eqref{DA} and \eqref{gauge_gauge}
imply $\cD_\al D^{\hat{a}} = -\ihalf \p_{\al\ad}
\bar{\lambda}^{\hat{a}\ad}$ and $\delta_\epsilon D^{\hat{a}}=0$). We
note that the Freedman-Townsend part \eqref{WZFT} can be directly
obtained from \eqref{sFT} by defining the component fields of $\Psi^A$
appropriately, but we skip the details of these definitions as they do
not matter.

The crucial part of the Lagrangian is \eqref{WZHK}. This part is neither
gauge invariant nor supersymmetric by itself. However, its gauge and
supersymmetry variations can be canceled (up to total derivatives) by
choosing the gauge and supersymmetry transformations of $B_{\mu\nu A}$,
$\varphi_A$ and $\chi_A$ appropriately, such that the gauge and
supersymmetry variations of \eqref{WZHK} are killed by terms in the
variations of \eqref{WZFT} (up to total derivatives). To show this, we
introduce the notation
 \begin{equation*}
  P = k_{ab}\, (\Ee^{-\Ii v} \lambda)^a (\Ee^{-\Ii v} \lambda)^b\ .
 \end{equation*}
Using the algebra \eqref{algebra}, one obtains straightforwardly
 \begin{equation} \label{calc1}
  \delta_a \cD^2 P = \frac{1}{4} \big[ - g^2 \Gamma^A \Gamma^B
  \T{A}{b}{c} \T{B}{c}{a} + 2g\, \Gamma^{\al A} \T{A}{b}{a} \cD_\al
  + g (\cD^\al \Gamma^A_{\!\al}) \T{A}{b}{a} - \delta^b_a \cD^2 \big]\,
  \delta_b P\ .
 \end{equation}
Similarly, the supersymmetry transformations of \eqref{WZHK} are
analysed, using \eqref{algebra} and the fact that $P$ is chiral,
$\cDb_\ad P=0$. The latter follows from the definition of
$\bar{\Gamma}^A_{\!\ad}$, \eqref{Gamma}, and from $\cDb_\ad
\lambda^a_\al$ given in \eqref{DA},
 \begin{equation} \label{chiral}
  \cDb_\ad (\Ee^{-\Ii v} \lambda_\al)^a = g \bar{\Gamma}^A_{\!\ad}
  (\Ee^{-\Ii v} T_A\, \lambda_\al)^a + (\Ee^{-\Ii v} \cDb_\ad\,
  \lambda_\al)^a = 0\ .
 \end{equation}
Using in addition $\cD_\al A_\mu^a$ given in \eqref{DA}, one finds
 \begin{align}
  \cD_\al \cD^2 P & = 0 \notag \\
  \cDb_\ad \cD^2 P & = 2\Ii\, \p_{\al\ad} \cD^\al P
  - \Ii \big[ 4 \bla_\ad^a - 2g\, \Gamma^{\al A} A^b_{\al\ad}
  \T{A}{a}{b} + 2 A_{\al\ad}^a \cD^\al \big]\, \delta_a P\ .
  \label{calc2}
 \end{align}
In order to analyse \eqref{calc1} and \eqref{calc2}, we use that
eq.\ \eqref{gauge_gauge} gives, due to the gauge invariance of $v$,
 \begin{equation} \label{calc3}
  \delta_a P = 2\, \Omega^\al_{Aa}\, \eta^A_\al\ ,
 \end{equation}
where 
 \begin{equation} \label{Om}
  \Omega_{\al Aa} = \la_\al^b G_{bc}(v) \T{A}{c}{a}\ ,\quad
  G_{ab}(v) = k_{cd}\, (\Ee^{-\Ii v})^c{}_a\, (\Ee^{-\Ii v})^d{}_b\ .
 \end{equation}
Eqs.\ \eqref{DV} show that each term in $\cD_\al \eta^A_\be$ and in
$\cD^2 \eta^A_\be$ contains exactly one of the fields $\eta^A_\al$,
$\bar{\eta}^A_\ad$, $h^A$ or $V_{\mu\nu}^A$, where all these fields
appear undifferentiated except for $\bar{\eta}^A_\ad$. From eqs.\
\eqref{calc1}, \eqref{calc2} and \eqref{calc3} we can thus infer that,
up to total derivatives, the gauge and supersymmetry variations of
\eqref{WZHK} can be written as linear combinations of the
undifferentiated fields $\eta^A_\al$, $\bar{\eta}^A_\ad$, $h^A$ and
$V_{\mu\nu}^A$ with field dependent coefficient functions. The
particular form of \eqref{WZFT} allows us therefore to cancel these
linear combinations through appropriately chosen terms in the
transformations of $B_{\mu\nu A}$, $\varphi_A$ and $\chi_A$ which are
obtained from evaluating \eqref{calc1} and \eqref{calc2} explicitly.

This yields the following gauge transformations of $ B_{\mu\nu A}$,
$\varphi_A$ and $\chi_A$,
 \begin{align} 
  \delta_\epsilon \varphi_A & = g \epsilon^a \delta_a \varphi_A\,
	\quad \delta_\epsilon B_{\mu\nu A} = g \epsilon^a \delta_a
	B_{\mu\nu A}\ ,\quad \delta_\epsilon \chi_A = g \epsilon^a
	\delta_a \chi_A \notag \\[4pt]
  \delta_a \varphi_A & = - \nabla \Omega_{Aa} - \bar{\nabla}
	\bar{\Omega}_{Aa} \notag \\[4pt]
  \delta_a B_{\mu\nu A} & = - \nabla \si_{\mu\nu} \Omega_{Aa} -
	\bar{\nabla} \bsi_{\mu\nu} \bar{\Omega}_{Aa} \notag \\[4pt]
  \delta_a \chi_{A\al} & = \half \nabla^2 \Omega_{\al Aa} + \Ii
	\p_{\al\ad} \bar{\Omega}^\ad_{Aa} + \Ii g\, V_{\al\ad}^B
	\bar{\Omega}^\ad_{Ca} \f{AB}{C}\ , \label{gauge_lin}
 \end{align}
where
\begin{align}
  \na{\al} \Omega_{\be Aa} & = \cD_\al \Omega_{\be Aa} - g
	\Gamma^B_{\!\al} (\T{B}{b}{a} \Omega_{\be Ab} + \f{BA}{C}
	\Omega_{\be Ca}) \notag \\
  \nabla^2 \Omega_{\be Aa} & = \cD^\al \na{\al} \Omega_{\be Aa}
	- g \Gamma^{\al B} (\T{B}{b}{a} \na{\al} \Omega_{\be Ab} 
	+ \f{BA}{C} \na{\al} \Omega_{\be Ca})\ . \label{naOm}
 \end{align}

Analogously one determines the terms in the supersymmetry
transformations of $B_{\mu\nu A}$, $\varphi_A$ and $\chi_A$ that
compensate for the supersymmetry variation of \eqref{WZHK}. The
supersymmetry transformations of $B_{\mu\nu A}$, $\varphi_A$ and
$\chi_A$ still have to be completed by contributions which cancel those
terms in the supersymmetry variation of \eqref{WZFT} originating from
the transformations of the auxiliary fields (up to total derivatives).
Not surprisingly, the additional contributions contain the standard
supersymmetry transformations of a linear multiplet, plus some nonlinear
extra terms involving the auxiliary fields. Altogether one finds
 \begin{align}
  \cD_\al \varphi_A & = \chi_{\al A} - \Ii g A^a_{\al\ad}
	\bar{\Omega}^\ad_{Aa} - g \Gamma^B_{\!\al} \f{AB}{C}
	\varphi_C \notag \\[4pt]
  \cD_\al B_{\mu\nu A} & = (\si_{\mu\nu} \chi_A)_\al + \Ii g A_\rho^a
	(\si^\rho \bsi_{\mu\nu} \bar{\Omega}_{Aa})_\al - g
	\Gamma^B_{\!\al} \f{AB}{C} B_{\mu\nu C} \notag \\[4pt]
  \cD_\al \chi_{\be A} & = - g \Gamma^B_{\!\al} \f{AB}{C}
	\chi_{\be C} \notag \\[4pt]
  \cDb_\ad \chi_{\al A} & = - \ihalf (\p_{\al\ad} \varphi_A
	+ g V_{\al\ad}^B \f{AB}{C} \varphi_C) - 2 \Ii g\,
	\bla_\ad^a\, \Omega_{\al Aa} - \Ii g A_{\be\ad}^a\,
	\nabla^\be\, \Omega_{\al Aa} \notag \\
  & \tab + \half \si_{\mu\al\ad}\, \ep^{\mu\nu\rho\si} (\p_{\nu}
	B_{\rho\si A} + g V_\nu^B \f{AB}{C} B_{\rho\si C}) - g
	\bar{\Gamma}^B_{\!\ad} \f{AB}{C} \chi_{\al C}\ .
	\label{DB}
 \end{align}

In addition the Lagrangian is gauge invariant under transformations
of $ B_{\mu\nu A}$ as in \eqref{gaugetrans}, with all other fields
invariant under these gauge transformations,
 \begin{equation} \label{WZgaugetrans}
  \delta_C B_{\mu\nu A} = \na{\mu} C_{\nu A} - \na{\nu} C_{\mu A}\
  ,\quad \delta_C\, (\text{all other fields}) = 0\ .
 \end{equation}

Let us now return to the algebra of supersymmetry and gauge
transformations. One finds that \eqref{gauge_lin} and \eqref{DB} realize
the algebra \eqref{algebra} on $B_{\mu\nu A}$, $\varphi_A$ and $\chi_A$
only on-shell\footnote{As in the nonsupersymmetric case,
$\com{\delta_a}{\delta_b}$ vanishes only on-shell.} and up to gauge
transformations \eqref{WZgaugetrans}. In particular
one gets
 \begin{align*}
  \aco{\cD_\al}{\cD_\be}B_{\mu\nu A} & = 0 \\
  \aco{\cD_\al}{\cDb_\ad}B_{\mu\nu A} & = - \Ii \cD_{\al\ad}
	B_{\mu\nu A} + \Ii (\na{\mu} Z_{\nu\al\ad A} - \na{\nu}
	Z_{\mu\al\ad A})\ ,
 \end{align*}
where
 \begin{equation*}
  Z_{\mu \nu A} = \half\, \eta_{\mu\nu}\, \varphi_A - B_{\mu\nu A}\
  ,\quad \na{\mu} Z_{\nu\rho A} = \p_\mu Z_{\nu\rho A} + g V_\mu^B
  \f{AB}{C} Z_{\nu\rho C}\ .
 \end{equation*}
Altogether, we find that the commutator of two supersymmetry
transformations involves a translation and gauge transformations
$\delta_\epsilon$ and $\delta_C$ with field dependent $\epsilon^a$ and
$C_{\mu A}$. More precisely, denoting a supersymmetry transformation
with anticommuting parameters $\xi$ by
 \begin{equation*}
  \Delta_\xi = \xi^\al \cD_\al + \bar{\xi}_\ad \cDb^\ad\ ,
 \end{equation*}
one gets on all the fields
 \begin{gather}
  \com{\Delta_\xi}{\Delta_{\xi'}} = a^\mu \p_\mu - \delta_\epsilon
	- \delta_C \notag \\
  a^\mu \equiv \Ii \xi' \sigma^\mu \bar{\xi} - \Ii \xi \sigma^\mu
	\bar{\xi'}\ ,\ \epsilon^a \equiv a^\mu A_\mu^a\ ,\
	C_{\mu A} \equiv \half a_\mu \varphi_A - a^\nu B_{\mu\nu A}\ .
 \end{gather}
In first order formulation this holds off-shell, in the formulation
without auxiliary fields only on-shell.

Finally, we spell out \eqref{WZHK} explicitly,
 \begin{align}
  L_\mathrm{HK} + L_\mathrm{CM} & = 
        G_{ab}(v) \big[ D^a D^b - \frac{1}{8}
	\hF_{\mu\nu}^a \hF^{\mu\nu b} + \frac{\Ii}{16}
	\ep^{\mu\nu\rho\si} \hF_{\mu\nu}^a \hF_{\rho\si}^b - \Ii \la^a
	\si^\mu \p_\mu \bla^b\, \big] \notag \\
  & \tab + \Ii g\, A_\mu^a \Omega_{Aa} \si^\mu \bar{\eta}^A - \Ii g\,
	V^A_\mu	\Omega_{Aa} \si^\mu \bla^a - \frac{g}{2} \big(
	\Omega_{Aa} + G_{ab}(v) \T{A}{b}{c} \la^c \big) \times \notag \\
  & \tab \times \big[ 2 \hat{\Gamma}^A D^a + \Ii \si^{\mu\nu}
	\hat{\Gamma}^A \hF_{\mu\nu}^a + g \T{B}{a}{d} \la^d\,
	\hat{\Gamma}^A \hat{\Gamma}^B\, \big] \notag \\
  & \tab + \frac{g}{2} G_{ab}(v) \big[ \Ee^{\Ii v} D^\al (\Ee^{-\Ii v}
	\hat{\Gamma}_{\!\al}^A T_A \Ee^{\Ii v})\Ee^{-\Ii v} \la \big]^a
	\la^b \notag \\
  & \tab + \text{c.c.}\label{fullHK}
 \end{align}
with the abbreviation
 \begin{equation}
  \hat{\Gamma}^A \equiv \Gamma^A + \Gamma^B (\Ee^{-\Ii \hat{v}}
  )_B{}^A\ .
 \end{equation}
{\sl Remarks.}

1. As $B_{\mu\nu A}$, $\varphi_A$ and $\chi_A$ appear only in the
Freedman-Townsend part $L_\mathrm{FT}$ of the Lagrangian, one
immediately concludes that $V_{\mu\nu}^A$, $h^A$ and $\eta^A$ vanish
on-shell. It is also to easy to infer that $h^A$ appears only linearly
in the action and that its equation of motion yields $v^A$ as a function
of the $\varphi_A$ (the precise relation between the $v^A$ and
$\varphi_A$ depends on the choice of the $T_A$ and the function
$\mathcal{F}$ in $L_\mathrm{aux}$).

2. The previous remark implies that the gauge
transformations of $\lambda^a$, $D^a$ and $\varphi_A$ vanish on-shell
(for $\lambda^a$ and $D^a$, this is seen from
\eqref{gauge_gauge1} because $\eta^A$ and $h^A$ vanish on-shell; for
$\varphi_A$, it follows from the fact that $\varphi_A$ equals
on-shell a function of the $v^A$). The algebra \eqref{algebra}
shows thus that, on these fields, the supersymmetry transformations
commute on-shell with all the gauge transformations. The same is
however not true for $A_\mu^a$ and $B_{\mu\nu A}$, as their gauge
transformations do not vanish on-shell.

3. As in usual supersymmetric gauge theories, a Fayet-Iliopoulos 
contribution breaks supersymmetry spontaneously, as is seen from the
equation of motion for $D^{\hat a}$ and from $\cD_\al
\lambda^{\hat a}_\be$ in \eqref{DA}. The gauge symmetries
remain unbroken, as one can infer from the fact that the gauge
transformation of $\varphi_A$ vanishes on-shell (cf.\ previous remark).

4. $g=0$ reproduces the usual supersymmetric gauge theories
for free real linear multiplets (in first order formulation)
and abelian WZ gauged vector multiplets. Hence, the models
are deformations of these standard supersymmetric gauge theories.
For $g\neq 0$ but $T_A=0$,
\eqref{fullHK}, \eqref{DA} and 
\eqref{gauge_gauge1} still reproduce the Lagrangian,
supersymmetry and gauge transformations of standard free abelian
supersymmetric gauge theory in WZ gauge (as $G_{ab}$ is
constant for $T_A=0$), while the linear and auxiliary
multiplets establish supersymmetric pure
Freedman-Townsend models in WZ gauge without
couplings to the abelian gauge multiplets 
$A_\mu^a$, $\lambda^a$, $D^a$.

\section{Examples} \label{example}

To illustrate some features of the models constructed in the previous
sections, we will now discuss two examples. We begin with the simplest
case of one gauge multiplet, one linear and one auxiliary multiplet. We
thus drop the indices $A$ and $a$ in the following, and take $T=1$. The
field dependent coupling $G$ and the spinor $\Omega$ defined in eq.\
\eqref{Om} reduce to
 \begin{equation}
  G(v) = k \Ee^{-2\Ii gv}\ ,\quad \Omega = k \Ee^{-2\Ii gv} \la\ ,
 \end{equation}
where we shall further simplify the discussion by considering $k=1$
only. In this case we have $\Gamma_{\!\al} = \Ii \cD_\al v$,
$V_{\al\ad}=\com{\cD_\al}{\cDb_\ad}v$, and the field strengths are
 \begin{equation}
  V_{\mu\nu} = \p_\mu V_\nu - \p_\nu V_\mu\ ,\quad \hF_{\mu\nu} =
  \p_\mu A_\nu - \p_\nu A_\mu + g(V_\mu A_\nu - V_\nu A_\mu)\ .
 \end{equation}
If we take $\mathcal{F}(V)$ to be quadratic, the complete Lagrangian
reads
 \begin{align}
  L_1 & = \frac{1}{2} V_\mu K^{\mu\nu} V_\nu - V_\mu \cH^\mu +
	\frac{1}{4} \p_\mu v \p^\mu v - \frac{\Ii}{2} \Gamma \si^\mu
	\overset{\leftrightarrow}{\p_\mu} \bar{\Gamma} + h (2v +
	\varphi) \notag \\
  & \tab + \eta ( \chi - 2\Ii\, \Gamma - \Ii g A_\mu\, \si^\mu
	\bar{\Omega}) + \bar{\eta} (\bar{\chi} + 2\Ii\, \bar{\Gamma}
	- \Ii g A_\mu\, \bsi^\mu \Omega) \notag \\
  & \tab + \frac{1}{2} (M + 2g\, \bar{\Omega} \bla)\, (\bar{M} + 2g\,
	\Omega \la) -2 g^2 \Omega \la\, \bar{\Omega} \bla + 2
	\cos(2gv) D^2 \notag \\
  & \tab  - \frac{1}{4} \cos(2gv)\, F_{\mu\nu} F^{\mu\nu} +
	\frac{1}{8} \sin(2gv)\, \ep^{\mu\nu\rho\si} F_{\mu\nu}
	F_{\rho\si} \notag \\
  & \tab - \Ii G(v) \la \si^\mu \p_\mu \bla + \Ii \bar{G}(v) \p_\mu \la
	\si^\mu \bla - 4g\, D (\Gamma \Omega + \bar{\Gamma}
	\bar{\Omega}) \notag \\[4pt]
  & \tab - 4 g^2 \Gamma \Gamma\, \la \Omega - 4 g^2 \bar{\Gamma}
	\bar{\Gamma}\, \bla \bar{\Omega} + 2\Ii g\, F_{\mu\nu}
	(\Gamma \si^{\mu\nu} \Omega + \bar{\Omega} \bsi^{\mu\nu}
	\bar{\Gamma})\ ,
 \end{align}
where $M=\Ii \cD^2 v$, and
 \begin{align}
  K^{\mu\nu} & = \eta^{\mu\nu} + g^2 \cos(2gv) (A^\mu\! A^\nu -
	\eta^{\mu\nu}\! A^\rho\! A_\rho) \\[4pt]
  \cH^\mu & = \half \ep^{\mu\nu\rho\si} \p_\nu B_{\rho\si} + \Ii
	g\, \Omega \si^\mu \bla - \Ii g\, \la \si^\mu \bar{\Omega}
	\notag \\
  & \tab + g \cos(2gv) F^{\mu\nu}\! A_\nu - g \sin(2gv)\,
	\ep^{\mu\nu\rho\si}\! A_\nu \p_\rho A_\si \notag \\
  & \tab - 4\Ii g^2 A_\nu\, (\Gamma \si^{\mu\nu} \Omega + \bar{\Omega}
	\bsi^{\mu\nu} \bar{\Gamma}) \\[4pt]
  F_{\mu\nu}&= \p_\mu A_\nu - \p_\nu A_\mu\ .
 \end{align}

Let us now discuss the formulation without auxiliary fields. By
virtue of the equation of motion for $h$ we can replace $v$ with
$-\half \varphi$. To eliminate $V_\mu$, we need to invert the matrix
$K^{\mu\nu}$. In the simple case at hand the inverse can be given
explicitly,
 \begin{equation}
  (K^{-1})_{\mu\nu} = \frac{1}{E}\, \big( \eta_{\mu\nu} - g^2
  \cos(g\varphi) A_\mu A_\nu \big)\ ,\quad E \equiv 1 - g^2
  \cos(g\varphi) A^\mu\! A_\mu\ .
 \end{equation}
We note that $\cH^\mu$ is of the form
 \begin{equation*}
  \cH^\mu = H^\mu + L^{\mu\nu}\! A_\nu\ ,\quad H^\mu \equiv \half
  \ep^{\mu\nu\rho\si} \p_\nu B_{\rho\si} - 2 g \sin(g\varphi)
  \la \si^\mu \bla
 \end{equation*}
with $L^{\mu\nu}$ antisymmetric. So the equation of motion for $V_\mu$
yields
 \begin{equation}
  \frac{1}{2} V_\mu K^{\mu\nu} V_\nu - V_\mu \cH^\mu \approx -
  \frac{1}{2E} \cH^\mu \cH_\mu + \frac{g^2}{2E} \cos(g\varphi)\,
  (A_\mu H^\mu)^2\ .
 \end{equation}
It proves convenient to eliminate $\chi$ in favor of $\Gamma$, which we
keep as an independent field instead. Variation with respect to $\eta$
then identifies $\chi$ as the combination
 \begin{equation*}
  \chi \approx 2\Ii\, \Gamma + \Ii g\, \Ee^{-\Ii g\varphi} A_\mu
  \si^\mu \bla\ .
 \end{equation*}
Elimination of $D$ gives rise to four fermion terms only, as a
Fayet-Iliopoulos term is not possible here,
 \begin{equation*}
  D \approx \frac{g}{\cos(g\varphi)}\, \big( \Ee^{\Ii g\varphi}
  \Gamma \la + \Ee^{-\Ii g\varphi} \bar{\Gamma} \bla \big)\ .
 \end{equation*}
Inserting the above expressions back into the Lagrangian, we finally
arrive at
 \begin{align}
  L_1 & \approx - \frac{1}{2E} \cH^\mu \cH_\mu + \frac{g^2}{2E}
	\cos(g\varphi)\, (A_\mu H^\mu)^2 + \frac{1}{16} \p_\mu \varphi
	\p^\mu \varphi - \frac{\Ii}{2} \Gamma \si^\mu
	\overset{\leftrightarrow}{\p_\mu} \bar{\Gamma} \notag \\
  & \tab - \frac{1}{4} \cos(g\varphi)\, F_{\mu\nu} F^{\mu\nu} -
	\frac{1}{8} \sin(g\varphi)\, \ep^{\mu\nu\rho\si} F_{\mu\nu}
	F_{\rho\si} \notag \\
  & \tab - \Ii \Ee^{\Ii g\varphi} \la \si^\mu \p_\mu \bla + \Ii
	\Ee^{-\Ii g\varphi} \p_\mu \la \si^\mu \bla - 2 g^2 \la \la\,
	\bla \bla \notag \\
  & \tab - \frac{2g^2}{\cos(g\varphi)} \big( \Ee^{\Ii g\varphi}
	\Gamma \la + \Ee^{-\Ii g\varphi} \bar{\Gamma} \bla \big)^2
	- 4 g^2 \Ee^{\Ii g\varphi} \Gamma \Gamma\, \la \la - 4 g^2
	\Ee^{-\Ii g\varphi} \bar{\Gamma} \bar{\Gamma}\, \bla \bla
	\notag \\
  & \tab + 2\Ii\, g\, F_{\mu\nu} (\Ee^{\Ii g\varphi} \Gamma
	\si^{\mu\nu} \la + \Ee^{-\Ii g\varphi} \bla \bsi^{\mu\nu}
	\bar{\Gamma})\ .
 \end{align}
\bigskip

As a second example, we present an $N=1$ supersymmetric counterpart of
the toy model in \cite{BD} and the $N=2$ supersymmetric models in
\cite{CdWFKST,DT}. In \cite{HK} it was observed that these
theories correspond to the case
 \begin{equation}
  \T{1}{a}{b} = \begin{pmatrix} 0 & 0 \\ 1 & 0 \end{pmatrix}\ ,
 \end{equation}
i.e., we now deal with two gauge multiplets, one linear and one
auxiliary multiplet. Again, as the index $A$ takes only
one value, we drop it in the following. Since we now get
 \begin{equation}
  \Ee^{-\Ii gvT} = \begin{pmatrix} 1 & 0 \\ -\Ii gv & 1
  \end{pmatrix}\ ,
 \end{equation}
the field dependent coupling reads
 \begin{equation}
  G_{ab}(v) = \begin{pmatrix} k_{11} - 2\Ii\, gv\, k_{12} - (gv)^2
  k_{22}\, & k_{12} - \Ii gv\, k_{22} \\ k_{12} - \Ii gv\, k_{22}
  & k_{22} \end{pmatrix}\ ,
 \end{equation}
with complex numbers $k_{ab}$. As the entries in the second column of
$T$ are zero, so is the second component of the doublet $\Omega_a$,
 \begin{equation}
  \Omega_a = \big(\, k_{22} \la^2 + (k_{12} - \Ii gv\, k_{22}) \la^1\,
  ,\, 0\, \big)\ ,
 \end{equation}
and it follows from eqs.\ \eqref{gauge_gauge}, \eqref{DV} and
\eqref{gauge_lin} together with \eqref{naOm} that the gauge
transformations with parameter $\epsilon^2$ act trivially on all
the fields, except for $A_\mu^2$ ($\delta_{\epsilon} A_\mu^2
= \p_\mu \epsilon^2+gV_\mu\epsilon^1$). The field strengths now are
 \begin{equation} \begin{split}
  \hF_{\mu\nu}^1 & = \p_\mu A_\nu^1 - \p_\nu A_\mu^1\ ,\quad V_{\mu\nu}
	= \p_\mu V_\nu - \p_\nu V_\mu\ , \\
  \hF_{\mu\nu}^2 & = \p_\mu A_\nu^2 - \p_\nu A_\mu^2 + g (V_\mu A_\nu^1
	- V_\nu A_\mu^1)\ .
 \end{split} \end{equation}

We shall again take $\mathcal{F}(V)$ to be quadratic. Due to the
increased complexity we give only the bosonic part of the Lagrangian,
 \begin{align}
  L_2 & = \frac{1}{2} V_\mu K^{\mu\nu} V_\nu - V_\mu \cH^\mu +
	\frac{1}{4} \p_\mu v \p^\mu v + h (2v + \varphi) + \frac{1}{2}
	M \bar{M} \notag \\
  & \tab + \re k_{22}\, \big( 2 D^2 D^2 - \frac{1}{4} F_{\mu\nu}^2
	F^{\mu\nu 2} \big) - \frac{1}{8} \im k_{22}\,
	\ep^{\mu\nu\rho\si} F_{\mu\nu}^2 F_{\rho\si}^2 \notag \\
  & \tab +  \big( \re k_{11} + 2gv \im k_{12} - (gv)^2 \re k_{22}
	\big)\, \big( 2 D^1 D^1 - \frac{1}{4} F_{\mu\nu}^1 F^{\mu\nu 1}
	\big) \notag \\
  & \tab - \frac{1}{8} \big( \im k_{11} - 2gv \re k_{12} - (gv)^2
	\im k_{22} \big)\, \ep^{\mu\nu\rho\si} F_{\mu\nu}^1
	F_{\rho\si}^1 \notag \\
  & \tab + 2 \big( \re k_{12} + gv \im k_{22} \big)\, \big( 2 D^1 D^2
	- \frac{1}{4} F_{\mu\nu}^1 F^{\mu\nu 2} \big) \notag \\
  & \tab - \frac{1}{4} \big( \im k_{12} - gv \re k_{22} \big)\,
	\ep^{\mu\nu\rho\si} F_{\mu\nu}^1 F_{\rho\si}^2 + \mu D^1
	\notag \\
  & \tab + \text{fermions}\ ,
 \end{align}
where
 \begin{align}
  K^{\mu\nu} & = \eta^{\mu\nu} + \half g^2 \re k_{22}\, ( A^{\mu 1}\!
	A^{\nu 1} - \eta^{\mu\nu}\! A^{\rho 1}\! A^1_\rho) \\[4pt]
  \cH^\mu & = \half \ep^{\mu\nu\rho\si} \p_\nu B_{\rho\si} \notag \\
  & \tab + g (\re k_{12} + gv \im k_{22})\, F^{\mu\nu 1}\! A_\nu^1
	+ g \im k_{22}\, \ep^{\mu\nu\rho\si}\! A_\nu^1 \p_\rho
	A_\si^2 \notag \\
  & \tab + g (\im k_{12} - gv \re k_{22})\, \ep^{\mu\nu\rho\si}\!
	A_\nu^1	\p_\rho A_\si^1 + g \re k_{22}\, F^{\mu\nu 2}\!
	A_\nu^1\ .
 \end{align}
As in this case the matrix $T$ has a vanishing first row, a
Fayet-Iliopoulos term has been added for $D^1$, spontaneously breaking
supersymmetry.

Elimination of the auxiliary vector $V_\mu$ works exactly as in the
previous example,
 \begin{equation} \begin{split}
  (K^{-1})_{\mu\nu} & = \frac{1}{E}\, \big( \eta_{\mu\nu} - \half g^2
  \re k_{22}\, A_\mu^1 A_\nu^1 \big)\ ,\quad E \equiv 1 - \half g^2
  \re k_{22}\, A^{\mu 1}\! A_\mu^1 \\
  \Rightarrow \quad & \frac{1}{2} V_\mu K^{\mu\nu} V_\nu - V_\mu
  \cH^\mu \approx - \frac{1}{2E} \cH^\mu \cH_\mu + \frac{g^2}{4E}\,
  \re k_{22}\, (A_\mu^1 H^\mu)^2\ .
 \end{split} \end{equation}
Comparing with the $N=2$ supersymmetric models \cite{CdWFKST,DT},
$A_\mu^2$ corresponds to the gauge field in the vector-tensor multiplet,
while $A_\mu^1$ is the analog of the vector field used to gauge the
central charge.

\section{Conclusion} \label{con}

We have constructed $N=1$ supersymmetric versions of all the models
presented in section \ref{HK}. The resulting supersymmetric models
are nontrivial deformations of the standard supersymmetric
gauge theories for free linear and vector multiplets. They have
several unusual properties as compared to other globally supersymmetric
gauge theories. We find particularly remarkable that, in the WZ gauge
constructed in section \ref{WZ}, the supersymmetry transformations do
not commute with all the gauge transformations, in contrast to the
formulation with linearly realized supersymmetry given in section
\ref{sfields}. We have presented arguments which suggest that this
unusual feature might be an inevitable property of this type of
supersymmetric models, but we admit that these arguments rely on our
construction and are therefore not completely cogent.

Another unusual feature of the WZ gauged models is that neither the
Henneaux-Knaepen nor the Chapline-Manton parts of the action are
supersymmetric by themselves but only together with the
Freedman-Townsend part, again in contrast to the formulation with
linearly realized supersymmetry. This property is less surprising
because, as already in the nonsupersymmetric case, 
the Henneaux-Knaepen and
Chapline-Manton parts of the action are not separately gauge invariant,
but only together with the Freedman-Townsend part.

Our results suggest several possible generalizations. For instance, one
may investigate extensions of the models constructed here by including
further fields. Furthermore, one might try to couple these models to
supergravity. Another interesting extension of our results would be
their generalization to $N=2$ supersymmetry. In particular this might
streamline and generalize the results of \cite{CdWFKST,DT}. A
possible starting point for such generalizations could be the algebra
\eqref{algebra} or suitably modified versions thereof.
\bigskip

\parindent 0em
\textbf{Acknowledgements} \\[6pt]
We thank Norbert Dragon and Sergei Kuzenko for stimulating and
enjoyable discussions. FB was supported by the Deut\-sche
For\-schungs\-ge\-mein\-schaft.
\parindent 1em

\begin{appendix}

\section{Appendix}

We use $\eta_{\mu\nu}= \mathrm{diag}(+,-,-,-)$ and
$\aco{D_\al}{\bD_\ad} = - \Ii \p_{\al\ad}$ (note the absence of a
factor 2 here). Apart from this, our conventions agree with those in
\cite{WB}. Supercovariant derivatives, mapping superfields into
superfields, are thus
 \begin{equation*}
  D_\al = \frac{\p}{\p \theta^\al} + \frac{\Ii}{2} (\si^\mu
  \bar{\theta})_\al\, \p_\mu\ ,\quad \bD_\ad = -\frac{\p}{\p
  \bar{\theta}^\ad} - \frac{\Ii}{2} (\theta \si^\mu)_\ad\,
  \p_\mu\ .
 \end{equation*}
The supersymmetry transformations of the component fields of a
superfield $\Sigma$ are related to the supercovariant derivatives of
$\Sigma$ through
 \begin{equation*}
  \cD_\al \Sigma = D_\al \Sigma\ ,\quad \cDb_\ad \Sigma = \bD_\ad
  \Sigma
 \end{equation*}
where $\cD_\al$ and $\cDb_\ad$ act only on the component fields
(and anticommute with the $\theta$'s).
\bigskip

The matrix valued superfields in \eqref{W} and \eqref{curved} are
related by
 \begin{equation*}
  \cW^A_\al\, T_A = W^A_\al \Ee^{\Ii V} T_A\, \Ee^{-\Ii V}\equiv
  \cW_\al\ .
 \end{equation*}
$\cW_\al$ satisfies a Bianchi identity familiar from super Yang-Mills
theory,
 \begin{equation} \label{bianchi}
  D^\al \cW_\al + g \aco{\tilde{\Gamma}^\al}{\cW_\al} - \text{c.c.}
  = 0\ ,\quad g \tilde{\Gamma}_{\!\al} = \Ee^{-\Ii V}\! D_\al\,
  \Ee^{\Ii V}\ .
 \end{equation}
\eqref{bianchi2} is equivalent to \eqref{bianchi}. This can be derived
from the identity
 \begin{equation} \label{useful}
  \Ee^{\Ii V}\, T_A\, \Ee^{-\Ii V} = (\Ee^{-\Ii \hV})_A{}^B\, T_B\ ,
 \end{equation}
which holds for any matrix representation $\{T_A\}$ of $\cG$ because
the entries of $T_A$ are $\cG$-invariant tensors. \eqref{useful}
implies
 \begin{equation*}
  \cW^A_\al = W_\al^B\, (\Ee^{-\Ii \hV})_B{}^A\ ,\quad
  \aco{\tilde{\Gamma}^\al}{\cW_\al} = - W_\al^B\,
  (\Ee^{-\Ii \hV})_B{}^A\, \com{\tilde{\Gamma}^\al}{T_A}\ .
 \end{equation*}
The commutator in the latter expression can be written as follows
 \begin{align*}
  g \com{\tilde{\Gamma}^\al}{T_A} & = \Ee^{-\Ii V} D^\al
	\{\Ee^{\Ii V} T_A\} - \{T_A\, \Ee^{-\Ii V}\}
	D^\al \Ee^{\Ii V} \\
  & = \Ee^{-\Ii V} D^\al \big[ (\Ee^{-\Ii \hV})_A{}^B T_B\,
	\Ee^{\Ii V} \big] - (\Ee^{-\Ii \hV})_A{}^B \Ee^{-\Ii V}
	T_B D^\al \Ee^{\Ii V} \\
  & = \{\Ee^{-\Ii V} T_B\, \Ee^{\Ii V}\}\, D^\al
	(\Ee^{-\Ii \hV})_A{}^B \\
  & = (\Ee^{\Ii \hV})_B{}^C T_C\, D^\al (\Ee^{-\Ii \hV})_A{}^B\ ,
 \end{align*}
where expressions $\{\dots\}$ have been rewritten using \eqref{useful}.
Altogether we get
 \begin{equation*}
  D^\al \cW_\al + g \aco{\tilde{\Gamma}^\al}{\cW_\al} =
  (\Ee^{\Ii\hV})_A{}^B\, D^\al (W_\al\, \Ee^{-2\Ii\hV})^A\, T_B\ ,
 \end{equation*}
which implies the equivalence of \eqref{bianchi2} and \eqref{bianchi}.

\end{appendix}

\end{document}